\def\flux{{\rm erg \, s^{-1} \, cm^{-2}}}

\def\etal{{et al. }}

\documentstyle[12pt,aaspp4,epsf]{article}

\begin{document}
\lefthead{Donahue \& Voit}
\righthead{$\Omega_m$ and the EMSS}
\slugcomment{Accepted to ApJ Letters, July 22, 1999}
\title{$\Omega_{m}$ from the Temperature-Redshift
Distribution of EMSS Clusters of Galaxies}
\author{Megan Donahue \& G. Mark Voit} 
\affil{Space Telescope Science Institute \\3700 San Martin Drive \\
Baltimore, MD 21218 \\ donahue@stsci.edu, voit@stsci.edu}

\begin{abstract}

We constrain $\Omega_{m}$ through a maximum likelihood analysis of temperatures and redshifts of 
the high-redshift clusters from the EMSS. 
We  simultaneously  fit  the low-redshift Markevitch (1998) sample
(an all-sky sample from ROSAT with $z=0.04-0.09$),
a moderate redshift EMSS sample from Henry (1997) 
(9 clusters with $z=0.3-0.4$), 
and a more distant EMSS  
sample (5 clusters with $z=0.5-0.83$ from Donahue et al. 1999) 
finding best-fit values of $\Omega_m = 0.45 \pm 0.1 $ for an open universe and 
$\Omega_m=0.27 \pm 0.1$ for a flat universe. 
 We individually  analyze the effects of our governing assumptions, 
including the evolution 
and dispersion of the cluster luminosity-temperature relation, the 
evolution  and dispersion of the cluster mass-temperature
relation, the choice of low-redshift cluster sample, and the accuracy
of the standard Press-Schechter formalism.  
We examine whether the existence of the massive distant cluster 
MS1054-0321 skews our results and find its effect to be small. 
From our maximum likelihood analysis we conclude that 
our results are not very sensitive to our
assumptions, and bootstrap analysis shows that our results are not
sensitive to the current temperature measurement uncertainties. The systematic 
uncertainties are $\sim \pm 0.1$, and  
$\Omega_m=1$ universes are ruled out at 
greater than 99.7\% ($3-\sigma$) confidence.

\end{abstract}

\section{Introduction}

Massive distant clusters  of galaxies can be used to constrain 
models
of cosmological 
structure formation (e.g. Peebles, Daly \&
Juszkiewicz 1989;  Arnaud \etal 1992; Oukbir \& Blanchard 1992; 
Eke, Cole \& Frenk 1996; Viana \& Liddle 1996; 
Bahcall, Fan \& Cen 1997; Donahue \etal 1998; 
Borgani \etal 1999). The mass function of clusters 
reflects the sizes and numbers of the original perturbations, and
the {\em evolution}
of the mass function depends sensitively on $\Omega_m$, the mean density
of matter. 
In a critical universe
with $\Omega_m=1$, perturbation growth continues forever, while in a low-density universe ($\Omega_m<1$),
growth significantly decelerates once $z \sim \Omega_m^{-1}-1$.

The Extended Medium Sensitivity Survey  (EMSS; Gioia \etal 1990; 
Henry \etal 1992) has proved cosmologically interesting because it
contains several massive high-redshift clusters (Henry 1997; Eke \etal 1998; 
Donahue \etal 99 -- hereafter D99). 
We report here our analysis of a complete, high-redshift sample
of clusters of galaxies culled from the EMSS, including the most distant EMSS
clusters (D99).  
We use maximum likelihood analysis to compare the unbinned 
temperature-redshift data with analytical
predictions of cluster evolution from Press-Schechter models, normalized
to two different low-z cluster samples (Henry \& Arnaud 1991; Markevitch 1998).  

Section 2  briefly describes the model for cluster evolution, 
\S3 describes the cluster samples, and
\S4 describes the
implementation of the maximum
likelihood technique.   Section 5 discusses our results and their sensitivity to various  assumptions, and 
\S6 outlines the results of a bootstrap resampling of our cluster catalogs. 
Section 7 summarizes our findings.

\section{The Model}

The Press-Schechter formula (Press \& Schechter 1974), as extended
by Lacey \& Cole (1993), adquately predicts the evolution 
of the cluster mass function ($dn/dM$) in numerical simulations (e.g.
Eke, Cole \& Frenk 1996 (ECF); Borgani \etal 1999; Bryan \& Norman 1998).
To obtain predicted cluster temperature functions ($dn/dT$) from
this mass function we use a mass-temperature ($M-T$) relation 
appropriate for all values of $\Omega_m$ (Voit \& Donahue 1998, 1999).
At $z=0$ we normalize this relation to the simulations of Evrard \etal (1996).
We will show in \S5 that the $M-T$ relation of ECF yields similar results.

In this description of cluster evolution, the three main variables are 
the mean density of the universe $\Omega_m$, the slope $n$ of the initial 
density perturbation spectrum near the cluster scale, and $\nu_{c}$, a
parameter that reflects the abundance of virialized perturbations on a 
given mass scale at a particular moment in time.
For a given $\Omega_m$, $n$, and $\nu_{c}$, the number of clusters per unit steradian 
expected in a given redshift range is 
$ (dn/dM) (dM/dT) (dV/dz) F(T,z) $ integrated 
over the relevant redshift and temperature ranges, where
 $F(T,z)$ is a window function
defined by the flux and redshift limits of a given sample and the luminosity-temperature
relation (See \S4).

\section{Cluster Samples}

Our fitting procedure compares three cluster samples 
each covering distinct redshift ranges 
to the model (\S2). 
The EMSS provided two samples of distant clusters. 
Because the EMSS has multiple flux limits (Henry 
\etal 1992), 
it is equivalent to multiple surveys each
with different flux limits and sky coverages. 
To compute the volumes associated with the EMSS samples, we  
correct the predicted flux of
a cluster to that measured within a $2.\arcmin4 \times 2.\arcmin4$ 
detection cell (Henry \etal 1992). 
The $z=0.5-0.9$ EMSS sample, described in D99, consists of 5 EMSS 
clusters at $z=0.5-0.9$ (Gioia \etal 1990; Henry \etal 1992.) 
These are all of the EMSS clusters with 0.5-3.5 keV fluxes 
$>f_x=1.33 \times 10^{-13} \flux$. 
(MS2053 may also belong in this
sample, but at a flux limit below what is listed in Henry \etal 1992.)
The $z=0.3-0.4$ sample is described in Henry (1997). D99 modified that
sample slightly by revising the redshift of one cluster upwards
to 0.54 (MS1241), leaving 9 clusters in 
the Henry sample. The Henry   
sample has already been
used to constrain $\Omega_m$ by Eke \etal (1998) and Henry (1997).
This paper extends the previous analysis from $z=0.4$
to $z=0.8$, at which the cluster evolution is expected to be much more dramatic.

To establish a baseline for assessing cluster evolution, we used two different
low-z samples: the Markevitch sample and the HEAO sample. 
The Markevitch sample of clusters from the ROSAT
All Sky Survey with $z=0.04-0.09$ (Markevitch 1998), 
covers a sky area of 8.23 steradians 
to a 0.2-2.5 keV flux limit of $2.0 \times 10^{-11} \flux$. 
The HEAO cluster sample
is also an all-sky sample (Henry \& Arnaud 1991), with a 2-10 keV 
flux limit of $3.0 \times 10^{-11} \flux$. To compute the volumes available to these samples 
we assume that the detection techniques in both cases were sensitive
to the total extended flux. 
We explore the consequences of our choice of low-redshift
sample in \S5. 

\section{Methods and Assumptions}

To assess how well cosmological models fit the cluster temperature 
data, we adopt the maximum likelihood technique described by 
Marshall \etal (1983). 
Specifically, we 
minimize the maximum likelihood function:
\begin{equation} 
S = -2 \sum_{i} [ \ln [\frac{dn}{dT}(z_i, T_i)] + \ln [\frac{dV}{dz}(z_i)] ]
+ 
2 \sum_{k} \int_0^\infty dT  
\int_{z_{min,k}}^{z_{max,k}(T)} 
\frac{dn}{dT} \frac{dV}{dz} \Omega_k F(T,z) dz  
\end{equation}
where $z_i$ and $T_i$ are the redshift and temperature of cluster $i$,
$z_{min,k}$ is the minimum redshift of sample $k$, and $z_{max,k}(T)$
is the maximum redshift at which a cluster  of temperature $T$ 
can be seen in sample $k$. 
$V$ is the comoving volume per unit solid angle, 
and $\Omega_k$ is the solid angle corresponding to sample $k$.
In practice, 
the temperature integral is calculated between
3 and 15 keV. Only clusters with temperature 
greater than 3 keV are included in the analysis. 
  Intervals
around the minimum $S$ are distributed like $\chi^2$ so differences 
in $S$ are 
similar to the familiar $\Delta \chi^2$.

The  $z_{max,k}(T)$ values for our samples depend on the cluster 
luminosity-temperature ($L-T$) relation. 
 Low redshift clusters of galaxies have
a fairly well-defined $L-T$  
relation (e.g. David et al. 1993; Markevitch
1998) that high-redshift clusters of galaxies seem to follow (D99; Mushotzky
\& Scharf 1998). This relationship has a finite dispersion which we
handle in two ways. One method 
is to replace the $L-T$ relation with a line bounding the lower 
$1-\sigma$ envelope in $L$ (Henry 1997),  explicitly 
compute $z_{max}(T)$ for each 
 flux limit, and set $F(T,z)=1$. 
The second method is to incorporate the dispersion
relation into the a window function $F(T,z)$ (Eke \etal 1998). We have 
done this calculation both ways, and both methods yield very similar 
results. Since the window function seems to be the most realistic description
of the data, we use it for our default analysis and the same dispersion as
assumed in Eke \etal (1998). 
We explore the effect of including evolution in the $L-T$ relation in \S5.

We vary 3 parameters for our model to span a cube of parameter space:
$0.1<\Omega_{m}<1.0$, spectral index $-2.8<n<-1.0$, and 
$2.4<\nu_{c0}<3.1$, where $\nu_{c0}=\nu_c(5~{\rm keV}, \,  z=0)$.
We compute a multidimensional matrix of $S$ 
for 25 temperatures between $T = 3$ to 15 keV.  
The 2-10 keV 
$L-T$ relation from David \etal (1993) 
defines the EMSS and HEAO volumes, appropriately
k-corrected. The volume of the 
Markevitch (1998) sample, our default low-z sample, 
is defined by its own $L-T$ relation. 

This procedure yields 
 a best fit of $\Omega_m=0.45\pm0.10$, $n=-2.4 \pm 0.2$, and
$\nu_{c0}=2.77^{+0.05}_{-0.09}$, corresponding to a $\sigma_8=0.64\pm0.04$ 
when $\Lambda=0$.  
We achieve a similar degree of correspondence between
observed and predicted temperature functions 
for flat models when $\Omega_m = 0.27 \pm 0.1$, 
$n=-2.2 \pm 0.2$, $\nu_{c0}= 2.62^{+0.08}_{-0.09}$, corresponding to   $\sigma_8=0.73^{+0.03}_{-0.05}$.
Figure 1 plots the observed temperature functions (D99)
 and the theoretical
temperature function corresponding to the 
best fit  to the temperature and redshift data for $\Lambda =0$. 
Note that we fit the discrete, unbinned
temperature-redshift data, not the binned temperature function.  
Figure 2 shows that our $3\sigma$ confidence limits
on $\Omega_m$ exclude $\Omega_m=1$.

Our value for the 
best-fit $\Omega_m$ is consistent with that derived for the
low-redshift subset of our data by Henry (1997), Eke \etal (1998),
and Viana \& Liddle (1999).  
However, Viana \& Liddle (1999) report less stringent constraints  
than the previous studies because they were more conservative about
uncertainties in the low-$z$ normalization of the temperature function. 
The maximum likelihood method we use naturally 
accounts for the uncertainty of the low-$z$ determination of the
normalization; we investigate the use of somewhat 
different low-$z$ samples in the next section. 
Because our sample extends to higher
redshifts, our 3$\sigma$ confidence limits are considerably
stronger than those found by earlier cluster studies.

\section{Results and Discussion}
Our best fit values for $\Omega_m$ are fairly robust.  
This section 
briefly describes the sensitivity of $\Omega_m$ to the assumptions in
our model and procedure. Results for various assumptions are listed
in Table 1.
 
\begin{enumerate}
\item {\bf Changing the low-redshift sample.}
If we use the
 updated HEAO sample (Henry \& Arnaud 1991, with
best-fit temperature updates provided by Henry, private communication)
instead of the Markevitch sample,  
 we obtain  $\Omega_m \sim 0.3$ rather than
$\sim 0.45$ because of the somewhat lower normalization at $z=0$. We
also used the Markevitch sample with uncorrected temperatures and a
flatter low-$z$ $L-T$ relation (Markevitch 1998), and obtained a 
somewhat lower best-fit value for $\Omega_m$, $\Omega_m=0.4 \pm 0.1$.

\item {\bf Varying the $M-T$ relation.} 
We find our bounds on $\Omega_m$ change little when we switch to the 
ECF (1996) $M-T$ relation. Because the best-fit $\Omega_m$ turns out to be 
$>0.3$, the unphysical behavior of the ECF $M-T$ relation at low $\Omega$
and low $z$ is not a factor. (See Voit \& Donahue 1999 for more details.)

\item {\bf Dispersion in the $M-T$ relation.} Our
default assumption was that the $M-T$ relation has a finite dispersion
of 7\% (Evrard \etal 1997). Neglecting the dispersion results in a
negligible difference in $\Omega_m$; increasing the dispersion to 20\% 
and using the ECF $M-T$ relation increases
$\Omega_m$ slightly to $0.50$. 
Dispersion in the $M-T$ relation scatters some of the more
numerous low-mass clusters to higher temperatures, making the 
observed temperature function flatter, and 
somewhat enhancing the observed numbers of hot clusters relative
to cool clusters.

\item {\bf Evolution of the $L-T$ relation.}  If we allow
the normalization of the $L-T$ relation to evolve such that $L \propto
T^{\alpha} (1+z)^A$ and $A=2$, we find no significant 
differences in $\Omega_m$ values.
This null result is in contrast to similar exercises in modelling 
the evolution of the cluster luminosity function (e.g. Borgani \etal
1999) or cluster number counts (Ebeling \etal 1999), where evolution of
the $L-T$ relation in the appropriate direction ($A\sim 2$) allows models
with larger $\Omega_m$ ($\sim 1$) to nearly fit. $L-T$ evolution only modestly 
affects the sample volume used to predict the distribution of cluster temperatures. 

\item {\bf Omitting MS1054-0321.} MS1054-0321, the
hottest ($kT=12.3$ keV) and most distant ($z=0.83$) 
cluster in our sample (Donahue \etal 1998), 
may well be anomalous.  However, omitting MS1054-0321 had virtually 
no effect on the
best fit $\Omega_m$.  

\item {\bf Missing high redshift clusters in the EMSS.} The EMSS
could be incomplete due to the use of a single detect cell aperture, 
which could bias its cluster selection in favor of high central
surface brightness even at high-z (Ebeling \etal 1999; Lewis \etal
1999). If the EMSS is missing clusters at higher redshift,
the values for $\Omega_m$  derived here are upper limits. 

\item {\bf Deviations from Press-Schechter orthodoxy.} Some numerical
simulations indicate that massive, high-z clusters might be more common than
the standard PS formula predicts (Governato \etal 1999; Evrard, private
communication).  
We have tested the effects of reducing the standard evolution
of $\nu_c$ by a factor $(1+z)^{-0.125}$ (Governato \etal 1999), and find
that the best fit $\Omega_m$ rises to $0.5\pm^{0.2}_{0.05}$. 
Of all the systematic effects, 
this one has the largest effect on the best fit $\Omega_m$. 
Even so, $\Omega_m=1$ is barely allowed at the $3\sigma$ level.

\item {\bf A larger high-redshift sample.} We
simulated the effect of tripling the size of the EMSS by
tripling the assumed sky coverage of the EMSS and replicating the 
existing $T-z$ data pairs. Tripling the
number of known clusters with $z=0.3-0.9$ reduces the statistical 
uncertainty of $\Omega_m$ by a factor of $\sim 2$. 
Because the uncertainty in the current estimate is now equal parts systematic
and statistical, theoretical refinements will be needed if we wish to take
full advantage of larger surveys.
\end{enumerate}

\section{Bootstrap Catalogs and Experimental Uncertainties}

In order to investigate the effects of measurement uncertainties 
within our cluster $T-z$ catalogs, 
we generated bootstrap catalogs with re-sampled temperatures.
For the EMSS clusters, we used the mean temperatures from Gaussian fits 
to  temperature probability distributions derived  from
the X-ray data (D99, Table 4).
Ten thousand boot-strap catalogs were  generated for each of the three
original samples, Markevitch (1998), Henry (1997), and D99. 
The number of clusters in each catalog was predetermined from a Poisson distribution
 based on the number of clusters in the original catalog.
Each set of data was then fit  
to obtain a best-fit $\Omega_m$, normalization,
and slope, using all of the standard assumptions. Out of 10,000 catalog combinations, we obtained a best fit $\Omega_m>0.95$ for only 3.
These three catalog combinations were the ones for which the low-redshift
catalog had a high number of clusters while the high-redshift cluster
catalogs were nearly empty. 
We got very similar results when we repeated  bootstrap re-sampling of the three catalogs while assuming temperature
measurement uncertainties for the EMSS clusters that were half the original
uncertainty. This similarity suggests that temperature measurement  
errors do not dominate the uncertainty in this method of estimating 
$\Omega_m$.

\section{Summary}

We have used a maximum likelihood Press-Schechter 
analysis of the temperatures and 
redshifts of two high-z  
EMSS samples of clusters of galaxies and two low-z all-sky samples
of clusters to constrain $\Omega_{m}$.
We find a simultaneous best fit to the low-z Markevitch (1998) sample,
a moderate-z EMSS sample from Henry (1997), 
and a high-z EMSS  
sample (D99)  
of $\Omega_m = 0.45 \pm 0.1 $ for an open universe and 
$\Omega_m=0.27 \pm 0.1$ for a flat universe, quoting statistical 
uncertainties only. 
Our results are not very sensitive to the assumptions 
within our cluster evolution model, 
with systematic uncertainties $\sim \pm 0.1$. 
Universes with $\Omega_m=1$ are ruled out at 
greater than 99.7\% ($3-\sigma$) confidence in the scenarios described here.

\acknowledgements
We acknowledge the NASA grants NAG5-3257, NAG5-6236, NAG5-3208
and NAG5-2570 for partial support of this work. We benefitted greatly 
from exchanges with J. Patrick Henry in the development of the 
code and by his generous release of revised temperature data for the
low-z Henry \& Arnaud (1991) sample.

\newpage

\begin{figure}
\plotone{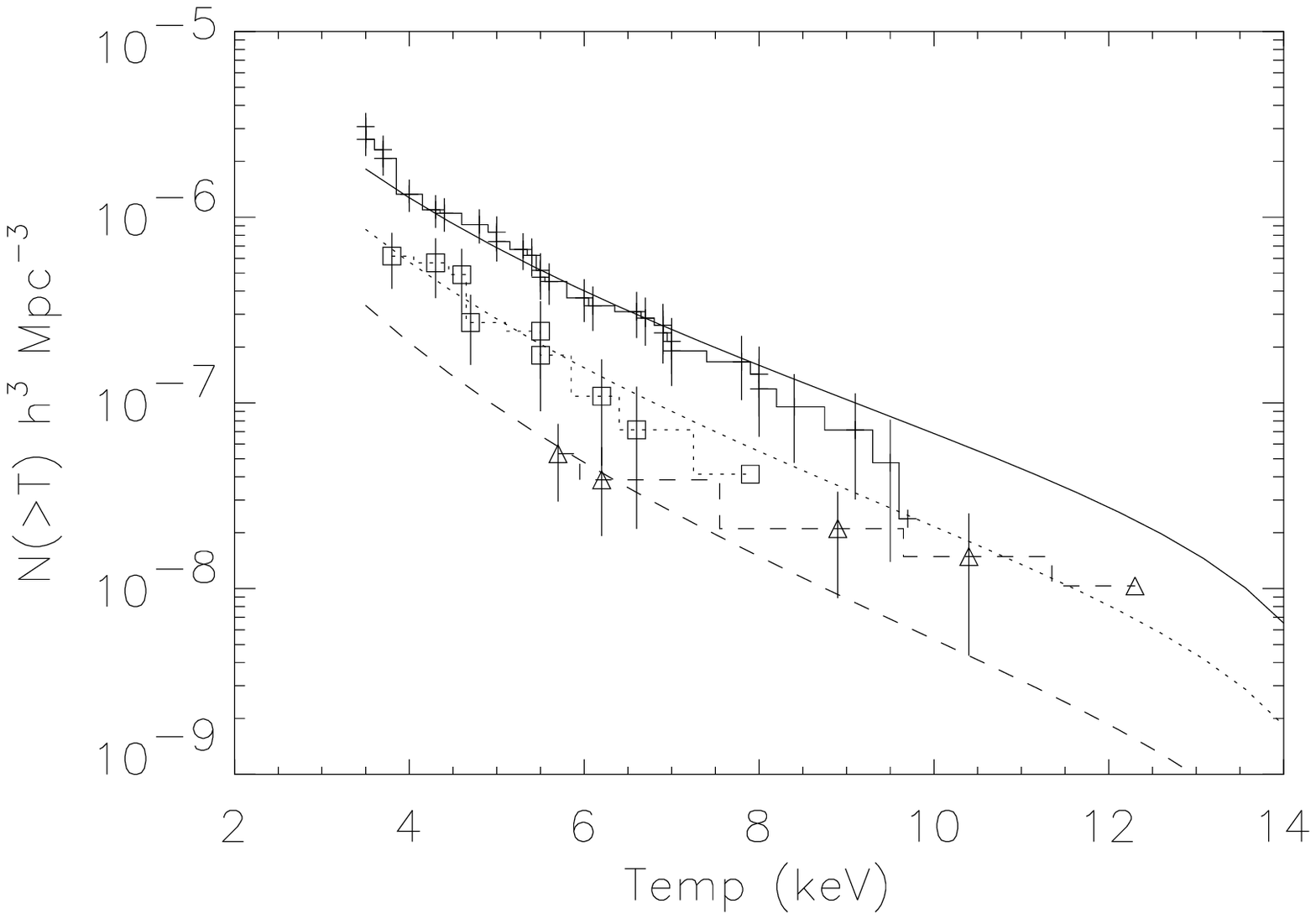}
\figcaption{Observed temperature functions for the clusters of galaxies in 
the Markevitch (1998) 
low-redshift sample (solid histogram) $z=0.04-0.09$, in the Henry (1997)
EMSS sample, $z=0.3-0.4$ (dotted histogram and squares), and in the
D99 EMSS sample $z=0.5-0.83$ (dashed histogram and triangles). The 
temperature function implied by the maximum likelihood best fit to the
temperature and redshift distribution in these three samples is overplotted
on each histogram as a smooth curve.}
\end{figure}

\begin{figure}
\plotone{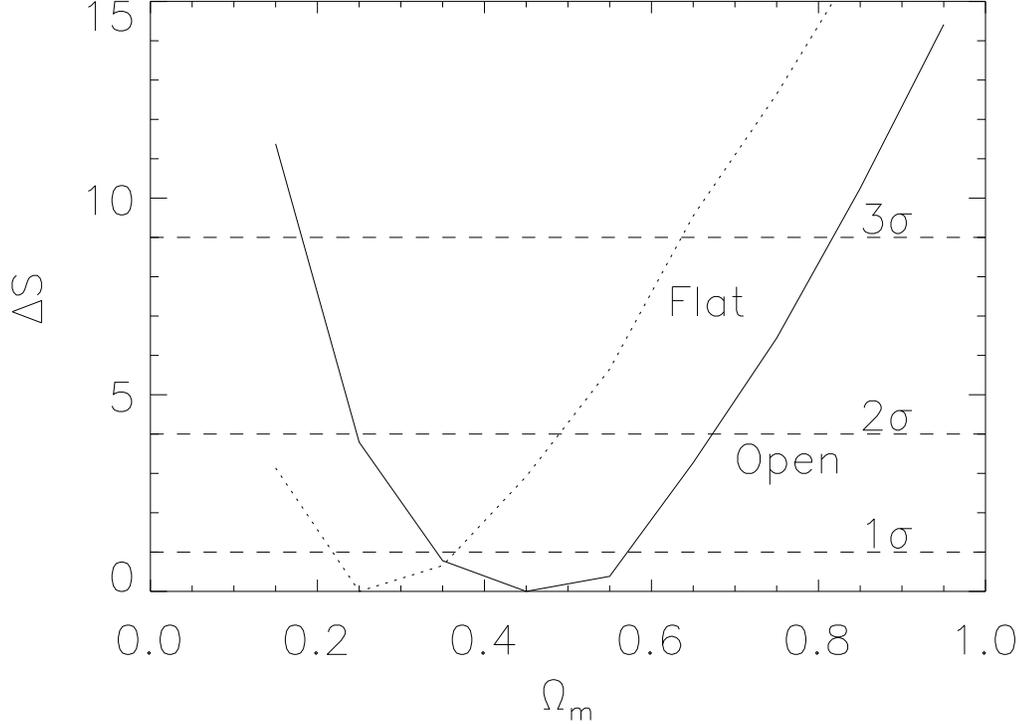}
\figcaption{Our distribution of $S$ for one interesting parameter, $\Omega_m$.
The 1, 2, and 3 $\sigma$ levels are marked in horizontal lines. The solid
line is for $\Lambda=0$ models and the dotted line is for a model with 
flat geometry ($\Lambda+\Omega_m =1$).}
\end{figure}

\newpage

\begin{table}
\caption{Results and Effects of Governing Assumptions}
\begin{tabular}{lcccc} \tableline
    & $\Omega$ & $\nu_c$ & $\sigma_8$ & $n$ \\ \tableline \tableline
Baseline Model & $0.45\pm0.1$ & $2.8\pm0.1$ & $0.64\pm0.04$ & $-2.3\pm0.2$ \\
Open, HEAO sub. for Markevitch &  $0.3\pm0.08$   & $2.9\pm0.1$ &$0.66\pm0.05$          	& $-2.0\pm0.2$      \\
Uncorrected Markevitch & $0.4\pm0.1$ & $2.8\pm0.15$ & $0.65\pm0.05$ 
& $-2.2^{+0.10}_{-0.25}$ \\
Flat, $\Lambda \neq 0$ & $0.27 \pm 0.1$ & $2.62\pm0.1$ & $0.73\pm0.05$ &
 $-2.2\pm0.2$  \\
No MS1054  & $0.5 \pm 0.1$ & $2.8 \pm ^{0.1}_{0.05}$ & $0.62 \pm 0.03$ & $-2.3\pm0.2$ \\
ECF M-T Relation  & $0.45\pm0.1$ & $2.8\pm0.1$ & $0.64\pm0.04$ & $-2.3\pm^{0.2}_{0.3}$ \\
L-T evolution $A=2$ & $0.45\pm0.1$ & $2.8\pm0.1$ & $0.62\pm0.04$ & $-2.3\pm0.2$  \\
No M-T dispersion & $0.45\pm0.1$ & $2.8\pm0.1$ & $0.64\pm0.04$ & $-2.3\pm^{0.2}_{0.3}$ \\
20\% M-T dispersion + ECF MT & $0.5\pm0.1$ & $2.8\pm^{0.05}_{0.10}$ & $0.64\pm0.04$ & $-2.3\pm0.2$ \\
Modified Press Schecter & $0.5 \pm0.1$ & $2.7\pm0.1$ & $0.63\pm0.03$ & $-2.5\pm0.15$ \\ \tableline
 \end{tabular}
\end{table}

\end{document}